\documentclass[preprint,11pt]{elsarticle}
\usepackage{lineno,hyperref}
\usepackage{natbib}
\usepackage{amssymb,amsmath}
\usepackage{algorithm,algpseudocode,algorithmicx,tabularx}
\usepackage{soul}
\usepackage{xcolor}
\usepackage[margin=1in]{geometry}
\usepackage{savesym}
\usepackage{amsmath}
\usepackage{url}

\usepackage{hyperref}
\hypersetup{breaklinks=true}
\usepackage{subcaption}
\savesymbol{iint}
\usepackage{txfonts}
\usepackage{multirow}
\usepackage{tikz}
\restoresymbol{TXF}{iint}
\usepackage{wrapfig}
\usepackage{graphicx}
\usepackage{url}
\usepackage{algorithm}
\usepackage{algpseudocode}
\modulolinenumbers[5]
\journal{Arxiv }

\begin{document}

\begin{frontmatter}

\title{Security and Privacy Enhancing in Blockchain-based IoT Environments via Anonym Auditing 
}

\author[rvt1]{Peyman Khordadpour}
\ead{peyman.khordadpor@yahoo.com}
\author[rvt]{Saeed Ahmadi}

\address[rvt1]{ 
Faculty of Computer Engineering,\\
University of Isfahan, Isfahan, Iran
} 

\address[rvt]{ 
Master of Cybersecurity and Threat Intelligence,\\
University of Guelph, ON, Canada
} 


\begin{abstract}
The integration of blockchain technology in Internet of Things (IoT) environments is a revolutionary step towards ensuring robust security and enhanced privacy. This paper delves into the unique challenges and solutions associated with securing blockchain-based IoT systems, with a specific focus on anonymous auditing to reinforce privacy and security. We propose a novel framework that combines the decentralized nature of blockchain with advanced security protocols tailored for IoT contexts. Central to our approach is the implementation of anonymization techniques in auditing processes, ensuring user privacy while maintaining the integrity and transparency of blockchain transactions. We outline the architecture of blockchain in IoT environments, emphasizing the workflow and specific security mechanisms employed. Additionally, we introduce a security protocol that integrates privacy-enhancing tools and anonymous auditing methods, including the use of advanced cryptographic techniques for anonymity. This study also includes a comparative analysis of our proposed framework against existing models in the domain. Our work aims to provide a comprehensive blueprint for enhancing security and privacy in blockchain-based IoT environments, paving the way for more secure and private digital ecosystems.
\end{abstract}

\end{frontmatter}

\section{Introduction}

The rapid evolution of the Internet of Things (IoT) and its integration with blockchain technology has opened new vistas in our digital world, particularly in enhancing security and privacy \cite{b1,a1}. This paper focuses on this emerging paradigm, exploring the interplay between blockchain technology and IoT environments. Our discussion centers on the unique security and privacy challenges that arise in blockchain-based IoT settings and how anonymous auditing can be a potent solution to these challenges \cite{a2,b2}. Blockchain's role as a foundational technology in IoT is crucial, providing a secure and transparent mechanism for handling the vast amounts of data generated by IoT devices \cite{a3}. This integration is essential for addressing security threats in IoT ecosystems, which are increasingly targeted due to their widespread use and inherent vulnerabilities \cite{b3,a4,a5}. The decentralized nature of blockchain, combined with advanced cryptographic techniques, offers a robust framework for securing IoT infrastructures against various cyber threats \cite{b4,b5}.

Industrial Internet of Things (IIoT) networks serve as a backbone for multiple applications, particularly in industrial environments like smart factories, enabling us to meet user demands \cite{a6,b6}. The benefits of blockchain technology have led to its widespread use in IIoT networks, including smart factories, homes, buildings, farms, cities, connected drones, and healthcare systems. This article primarily addresses blockchain-based IIoT network security in smart factories, but the proposed framework can be applied to other IIoT environments as well. In contemporary smart factories, numerous devices connect to public networks, and a variety of operations rely on smart systems such as temperature monitoring, Internet-connected lights, IP cameras, and phones. These devices handle private, sensitive data and may provide critical safety services\cite{a7}. As the number of IIoT devices in smart factories grows, securely storing, collecting, and sharing data becomes a major concern. Hence, industrial, critical, and personal data are at risk. Blockchain technology can maintain data integrity within and beyond smart factories by providing strong authentication and ensuring the availability of communication networks. However, privacy and security remain significant issues in IIoT networks, with the risk of fraudulent activities in blockchain-based systems\cite{b7,b8}. Smart factories must safeguard user data privacy throughout its transmission, usage, and storage. Data stored in these systems are susceptible to tampering by malicious actors aiming to access, modify, or misuse it. These attacks are often statistical outliers, showing significant deviations from normal behavior. Identifying such out-of-norm events is crucial for threat-hunting initiatives and securing systems against unauthorized access. This involves the automatic detection and filtering of anomalous activities to maintain system integrity and security \cite{a8,a9}.

However, while blockchain brings numerous security benefits, it also introduces new complexities, particularly regarding user privacy. In IoT contexts, where user data is often sensitive, ensuring anonymity and data privacy becomes paramount \cite{b9,b10}. This paper proposes a novel security protocol for blockchain-based IoT environments that emphasizes privacy enhancement and anonymous auditing. Our approach leverages the inherent security features of blockchain while integrating privacy-preserving techniques to protect user identities and data \cite{a10,a11}. We also explore the vulnerabilities inherent in smart contracts within IoT applications, highlighting the need for robust security protocols to mitigate these risks \cite{b11,a12}. The paper discusses various existing blockchain platforms and protocols that focus on privacy and security, providing a context for our proposed solution \cite{b12,b13}.

Our motivation is to design a security protocol that not only leverages the strengths of blockchain technology in IoT environments but also addresses the critical need for privacy and anonymity \cite{a14}. The protocol aims to ensure a secure, private, and auditable framework for IoT interactions, enhancing trust and reliability in these increasingly prevalent digital ecosystems \cite{b15,b14}. Our goal is to contribute to the secure advancement of IoT through blockchain technology, offering a comprehensive solution for enhancing both security and privacy in this rapidly evolving domain \cite{a13,a14}. The growing interest in blockchain technology's integration within Internet of Things (IoT) environments has spurred numerous discussions and research efforts. However, similar to the discourse surrounding Web 3.0, there is a lack of precise definitions and concrete architectural designs specifically addressing the unique requirements of blockchain in IoT contexts. Most existing studies tend to focus on high-level concepts or certain aspects like consensus algorithms, leaving other critical components and architectural designs underexplored. This gap in research indicates that the full potential of blockchain in IoT is either not fully realized or is branching into multiple directions of development. This paper aims to provide a focused analysis of blockchain architecture in IoT environments, highlighting the specific challenges and solutions in this domain. The following contributions are made in this paper:

\begin{itemize}
\item Design and present an architecture for integrating blockchain technology within IoT environments, detailing the workflow and specific security mechanisms that are essential for IoT applications.
\item Propose a security protocol tailored for blockchain-based IoT systems, emphasizing the use of privacy-preserving techniques and anonymous auditing during runtime to enhance security and privacy.
\item Discuss the application of privacy-enhancing techniques specifically suited for IoT environments.
\item Explore the use of anonymization methods, such as Tor, for conducting anonymous audits in IoT settings, ensuring both transparency and privacy.
\end{itemize}

The remainder of the paper is organized as follows: Section 2 delves into related work on blockchain technology in the context of IoT. Section 3 describes the proposed architecture for a blockchain-based IoT system. Section 4 focuses on the implementation of anonymous auditing within this architecture. Section 5 discusses the evolution and future prospects of blockchain technology in IoT environments. Section 6 examines the integration of blockchain with new and emerging technologies in IoT. The paper concludes in Section 7 with a summary of findings and suggestions for future research directions.

\section{Related work}\label{ss}
Integrating blockchain technology into IoT is essential due to the centralized nature of IoT systems, which typically involve a third-party authority controlling all data with unclear usage restrictions (Da Xu et al., 2021). In contrast, blockchain offers a decentralized, autonomous, trustless, and distributed environment. It addresses the centralization issues like single points of failure, trust, and security by leveraging the processing power of all participants, enhancing efficiency and removing single points of failure. Additionally, blockchain ensures superior security and data integrity through its immutable characteristics.

In terms of IoT security, based on blockchain networks, there are two primary threat types: privacy threats at the blockchain layer and security threats at the IoT layer \cite{b16}. Network attacks encompass both these threats, including Denial of Service (DoS), probing, Remote to Local (R2L), transaction privacy leakage, and phishing  \cite{b17}. Moreover, IoT networks face challenges with data uncertainty and vagueness, exacerbated by the heterogeneous nature of IoT data sources  \cite{a15}. While Deep Learning (DL) models are effective in processing cybersecurity dataset information and combatting attacks, they struggle with managing uncertainty, highlighting the need for more robust security solutions in blockchain-based IoT networks.

The intersection of IoT and blockchain is a critical area of research, focusing on addressing security threats more effectively. The paper argues for the importance of practical security approaches like efficient threat detection in these networks. The authors propose a novel approach: a secure intelligent fuzzy blockchain framework that leverages deep learning and fuzzy logic for improved threat detection in IoT networks. DL is integral in cybersecurity strategies for its ability to analyze attack patterns and adapt to changing behaviors, while fuzzy logic aids in making reasoned decisions under uncertain conditions \cite{b18}. This combination is poised to effectively address cyber threats and bring flexibility in decision-making and transaction acceptance in blockchain-based IoT networks \cite{a16}.

Research in blockchain technology, particularly in the Blockchain Edge of Things (BEoT), indicates its potential to enable future services and applications. This area explores how blockchain extends beyond its initial cryptocurrency use, offering benefits like decentralization, immutability, and traceability in Edge of Things (EoT) systems \cite{b19}.

Another study highlighted the application of Federated Learning (FL) for anomaly detection in smart buildings. This approach, which includes a recurrent neural network, emphasizes privacy and has proven to be over twice as fast in training compared to centralized methods. It also showed enhanced performance in classification and regression tasks \cite{a17,a18}.

Nguyen et al. introduced a self-learning federated system for identifying anomalies in IoT networks. This system, one of the first of its kind, relies on device communication profiles to spot unusual changes in IoT devices' communication patterns, using FL to efficiently aggregate these profiles. It's capable of addressing a broad range of emerging threats\cite{b20}.

Additionally, there was an approach using FL to detect abnormal client behavior at the server level. This method utilized low-dimensional representations of model weight vectors to identify network anomalies, significantly outperforming traditional defense-based methods\cite{b21}.

Furthermore, research involving Deep Learning and blockchain-based FL focused on detecting COVID-19. This approach involved collecting data from various sources to create a global deep learning model. Blockchain was used to validate data, ensuring privacy during the federated training process. This combination of blockchain and federated e-learning enabled the collaborative training of global models, demonstrating improved performance in patient detection\cite{a19,a20}.

\section{Blockchain-based IoT System Architecture}

The architecture for a blockchain-based Internet of Things (IoT) system is designed with openness and adaptability in mind, allowing for seamless integration across various devices and systems. It supports the development of user-generated content, interoperability among diverse IoT devices, and efficient communication protocols. This architecture is particularly tailored to address the unique challenges and opportunities in blockchain-enabled IoT environments. In this paper, we introduce a novel system architecture for blockchain-based IoT, as illustrated in Figure 1. This architecture facilitates interactions between users, IoT devices, and blockchain infrastructure, enabling secure and efficient delivery of IoT services\cite{b22}.

Figure 1 delineates the workflow and structural components of the proposed system. The core of this architecture lies in its decentralized approach, where data processing and storage occur in a distributed network controlled by community-driven protocols, a departure from traditional centralized Trusted Third Parties (TTPs). A distinctive aspect of this blockchain-based IoT system is the integration of immediate rewards mechanisms, ensuring equitable compensation for network participants contributing resources or data\cite{a21,a22}. The proposed system architecture is composed of four key layers: blockchain, application, client, and wallet. The blockchain layer is the backbone, providing a persistent ledger and smart contract functionality for managing on-chain data. It supports various distributed storage solutions like IPFS (InterPlanetary File System) and Swarm, incentivizing users for their distributed storage contributions. This layer accommodates diverse types of blockchains, including single, homogeneous, and heterogeneous networks.

At the application layer, we implement the necessary business logic and interactions for IoT devices. This layer manages data that doesn’t require blockchain storage, enabling efficient operation of various IoT applications. It also includes technologies for dynamically enhancing the underlying blockchain, improving overall system usability. The client layer offers interfaces for user interaction with IoT devices and blockchain services. Users connect to the blockchain through wallets embedded in browsers or applications, enabling secure transaction signing and management. In scenarios with high request volumes, an agent-based approach is employed to manage the load, ensuring smooth and scalable system performance. Wallets play a crucial role in the user experience, providing intuitive access to blockchain-based IoT services. Users can integrate wallets into their browsers or IoT applications, facilitating direct interaction with the blockchain. Wallets support various functions like transaction signing and management, crucial for secure and efficient IoT operations.
Security protocols are vital for ensuring the integrity and confidentiality of data within the blockchain-based IoT system. These protocols establish rules for secure data transfer, authentication, and encryption, protecting the system from unauthorized access and cyber threats. Without such protocols, the security of IoT networks would be significantly compromised.
In the subsequent sections, we will delve deeper into the specific roles and functions of each layer and entity in our blockchain-based IoT system, emphasizing their contributions to the overall security and efficiency of the architecture.

 \begin{figure}[H]
    \centering
    \includegraphics[width=0.67\linewidth,keepaspectratio]{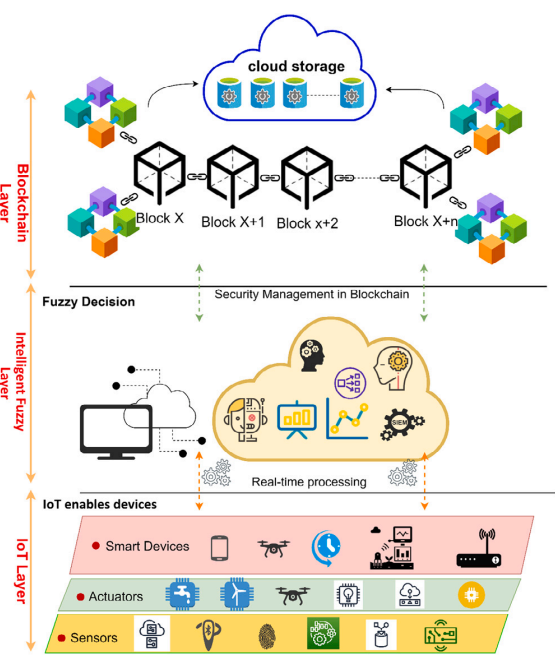}
    \caption{Workflow of IoT in Fuzzy Blockchain System}
    \label{fig11}
\end{figure}

\subsection{Privacy Enhancement in Blockchain-based IoT Systems}

In the realm of blockchain-based IoT systems, privacy enhancement is not just a feature but a necessity. The integration of blockchain technology in IoT environments offers a new paradigm for data management, where users have greater control over their data through private and public keys, ensuring ownership and preventing unauthorized access by third parties. When personal information is stored on a blockchain within an IoT system, users can set conditions for third-party access, bolstering data privacy and security. One innovative approach to privacy in blockchain-based IoT systems is the use of decentralized identity mechanisms. These systems leverage blockchain to enable individuals and organizations to manage their digital identities autonomously, contrasting the centralized identity systems controlled by singular entities like governments or corporations\cite{b22,b23}.

For instance, the concept of an Identity Hub in blockchain-based IoT ecosystems allows users to store and encrypt personal information such as names and addresses on a decentralized platform. This data is secured on the blockchain, ensuring immutability and security. Users can then selectively share their information with third parties using identity credentials, such as digital certificates or blockchain tokens, thus confirming their identity without divulging all personal details. Moreover, blockchain-based IoT systems can harness zero-knowledge proofs to enhance privacy. These proofs enable a user (the prover) to demonstrate possession of certain information to another party (the verifier) without revealing the information itself. For example, a user might need to prove that they meet a certain age requirement to access a service without disclosing their actual age\cite{a23}.

Self-sovereign identity is another pivotal aspect of privacy in blockchain-based IoT environments. This concept underscores the individual's right to full control and ownership of their digital identity and personal data. Self-sovereign identity systems in blockchain-based IoT enable individuals to store their personal data on a decentralized platform, managing access through identity credentials and smart contracts.
 \begin{figure}[H]
    \centering
    \includegraphics[width=0.67\linewidth,keepaspectratio]{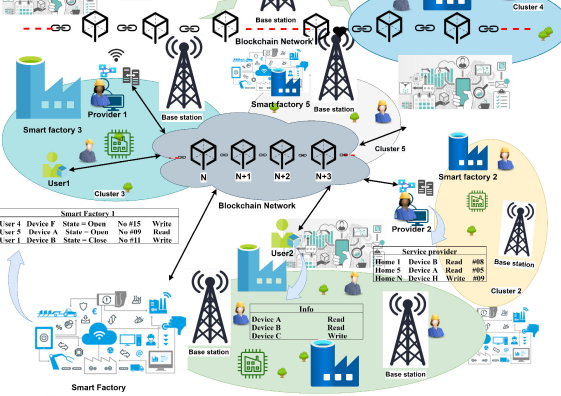}
    \caption{Workflow of IoT-Blockchain System}
    \label{fig11}
\end{figure}
\subsection{Formulating Privacy in Blockchain-based IoT Environments}

Privacy in blockchain-based IoT environments is a multi-faceted concept, encompassing various dimensions such as information privacy, physical privacy, and decisional privacy. Information Privacy: In blockchain-based IoT systems, information privacy involves the right of individuals to control the collection and use of their personal data. This includes both Personally Identifiable Information (PII) and non-PII. Blockchain technology can enforce principles like data minimization and purpose limitation, ensuring that only necessary data is collected for specific, legitimate purposes \cite{b24,a24}.

Physical Privacy: This dimension relates to the individual's right to maintain their physical space and personal property free from intrusion. In the context of IoT, this could involve ensuring that IoT devices respect user privacy and do not intrude upon personal spaces without consent. Decisional Privacy: Pertaining to the right to make personal choices without external interference, decisional privacy in a blockchain-based IoT environment could mean allowing users to make informed decisions about their data and how it is used. Key principles such as data minimization, purpose limitation, storage limitation, accuracy, integrity, and confidentiality are crucial in formulating privacy in these environments. Blockchain and IoT technologies together can provide mechanisms for ensuring data accuracy, integrity, and confidentiality, thereby protecting against unauthorized processing and potential data breaches  \cite{y1,a25,b25}.

In summary, the integration of privacy-enhancing technologies in blockchain-based IoT systems offers significant potential for enhancing online privacy and security. By adopting decentralized identity systems, zero-knowledge proofs, and self-sovereign identity concepts, these systems can provide users with a more secure, private, and controlled digital experience, ensuring the integrity and confidentiality of personal data.
 \begin{figure}[H]
    \centering
    \includegraphics[width=0.67\linewidth,keepaspectratio]{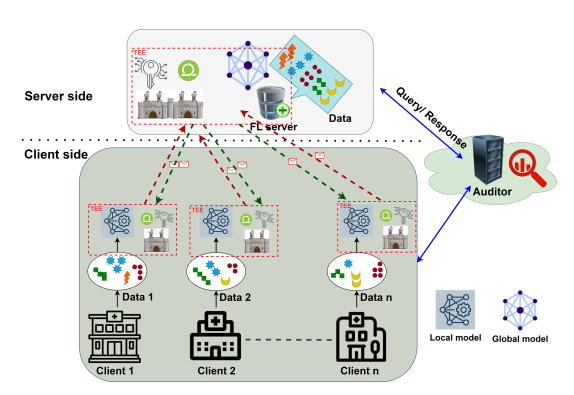}
    \caption{Workflow of A FL-IoT System}
    \label{fig11}
\end{figure}

\subsection{Privacy Preservation in Blockchain-based IoT Systems}

Privacy preservation is of paramount importance in the context of blockchain-based IoT systems, where the protection of personal and sensitive data is crucial in an environment of constantly expanding digital information. Privacy preserving techniques in these systems are designed to shield personal data from unauthorized access while still enabling valuable computations and analyses to be performed on the data. One key technique in privacy preservation is differential privacy, which offers a mathematical framework to assess privacy loss. By introducing noise into data query results, differential privacy ensures that the inclusion or exclusion of an individual's data does not significantly affect the overall outcome, thereby offering robust privacy guarantees\cite{a25,a26}.
Other significant privacy preservation methods include homomorphic encryption and secure multi-party computation. Homomorphic encryption allows computations to be carried out on encrypted data without needing to decrypt it first, maintaining data confidentiality throughout the process. Secure multi-party computation enables multiple entities to jointly compute a function over their inputs while keeping these inputs confidential from each other. Anonymization and pseudonymization are also crucial strategies in privacy preservation within blockchain-based IoT systems. Anonymization involves completely stripping personally identifiable information from datasets, rendering individual identification impossible. Pseudonymization, in contrast, involves replacing personal identifiers with pseudonyms, allowing individual identification under specific conditions but generally preserving privacy \cite{z7}.

\section{Anonymous Auditing in Blockchain-based IoT Systems}

Anonymous auditing is a critical component in ensuring compliance with privacy standards and maintaining confidentiality in blockchain-based IoT systems. It involves inspecting and evaluating systems or data sets to verify adherence to privacy regulations, with a focus on preserving the anonymity of individuals involved.
This process incorporates privacy-preserving techniques to ensure that the identities of individuals are not disclosed during audits. Methods like k-anonymity, l-diversity, and t-closeness are employed to maintain individual anonymity within datasets while still providing insightful data for auditing purposes. K-anonymity ensures that each individual in a released dataset is indistinguishable from at least k-1 others, protecting their identity. L-diversity enhances this by ensuring diverse and representative sensitive attributes within each group of k-indistinguishable individuals. T-closeness extends these concepts by requiring that the distribution of a sensitive attribute in any subset of the data closely resembles its distribution across the entire dataset\cite{z1}.

\subsection{Formulating Auditing in Blockchain-based IoT Systems}

In a blockchain-based IoT context, auditing is a systematic, objective assurance and consulting activity designed to add value and improve an organization's operations. This activity helps organizations achieve objectives through a disciplined approach to evaluate and improve the effectiveness of risk management, control, and governance processes. Key principles of auditing in this context include independence, evidence-based reporting, and relevance. Auditors must maintain objectivity and distance from the systems, processes, or organizations they audit to ensure impartial assessments. Their findings and conclusions should be based on evidence gathered during the audit process, enhancing the reliability and validity of their reports. Audits should also focus on relevant aspects aligned with the objectives and scope of the audit, aiming to improve operational efficiency \cite{z2}.

Auditing methodologies in blockchain-based IoT environments may involve various tools and techniques, including control self-assessment, risk assessment, and benchmarking. Advanced technologies like artificial intelligence and machine learning are increasingly employed for data analysis, enabling auditors to sift through vast datasets, detect anomalies, and identify trends.
Privacy-related auditing focuses on compliance with laws and regulations such as GDPR and CCPA. Auditors assess the effectiveness of an organization's data protection controls, including privacy policies, procedures, and technical measures implemented to safeguard personal data. This comprehensive evaluation ensures that blockchain-based IoT systems not only function efficiently but also adhere strictly to evolving privacy standards and regulations\cite{z3}.

\subsection{Auditing in Blockchain-based IoT Systems}

In the realm of blockchain-based IoT systems, auditing extends beyond traditional financial reviews to include the examination of smart contract execution, compliance with decentralized governance protocols, and ensuring data integrity within IoT applications. The inherent transparency and immutability of blockchain technology make it an effective tool for auditing in these environments.
Auditing in blockchain-based IoT systems can offer real-time monitoring and verification of transactions and activities, a significant advancement over traditional auditing methods. Due to the public nature of blockchain transactions, various stakeholders, including auditors, regulators, and users, can continuously monitor and verify activities on the blockchain. Moreover, the integration of advanced cryptographic tools, such as zero-knowledge proofs, enhances the auditing process by providing assurances of transaction integrity and authenticity without compromising privacy. This feature is particularly valuable in blockchain-based IoT systems, where sensitive data often needs to be protected from exposure\cite{a27,y2,a28}.

In addition to transaction verification, auditing in these systems supports decentralized governance models. Many IoT platforms leveraging blockchain technology implement governance protocols that allow network participants to make decisions collectively. In such models, auditing ensures the legitimacy of votes and adherence to the established protocols. Various tools and frameworks have been developed to facilitate auditing in blockchain-based IoT systems. Blockchain explorers enable users to track transactions, while smart contract analysis tools help detect vulnerabilities and verify contract behaviors. Furthermore, the emergence of decentralized auditing platforms offers independent auditing services for these systems, ensuring integrity and compliance with privacy standards \cite{y2}. However, challenges exist, including the technical complexity of blockchain and smart contracts, which may pose difficulties for those without specialized knowledge. Additionally, the transparency feature of blockchain raises privacy concerns, necessitating the integration of privacy-preserving technologies like zero-knowledge proofs to balance transparency with privacy\cite{z4,z5}. As blockchain-based IoT systems continue to evolve, the role of auditing becomes increasingly critical, not only for ensuring integrity and compliance but also for supporting decentralized governance and fostering wider adoption of these technologies.

\section{Evolution of Blockchain-based IoT Systems}

The evolution of blockchain-based IoT systems signifies a major shift in digital interactions, emphasizing semantic data interoperability and collaborative networks. These systems are evolving to address new paradigms of privacy and anonymity, especially in blockchain environments. Privacy has become a foundational necessity in blockchain-based IoT systems. The decentralization feature of blockchain technology shifts data handling from centralized repositories to peer-to-peer networks. This change demands robust privacy-enhancing techniques to protect sensitive user information without compromising the collaborative nature of these systems. Techniques like differential privacy, homomorphic encryption, and secure multi-party computation have become crucial. Differential privacy injects noise into data queries to ensure minimal impact from individual data, while homomorphic encryption allows computations on encrypted data, mitigating privacy breach risks. Secure multi-party computation enables collaborative data analysis while preserving individual privacy\cite{a28,a29}.

The transparency of blockchain transactions in IoT systems promotes trust and accountability. However, this transparency can infringe on privacy if identities are linked to transactions. Thus, anonymity is essential, leading to the development of anonymized auditing methodologies. Anonymous auditing in blockchain-based IoT systems involves validating transaction integrity and compliance with privacy standards while keeping participant identities hidden. Techniques like k-anonymity, l-diversity, and t-closeness provide statistical privacy guarantees, ensuring sensitive attributes are well-represented and maintain similar distributions across datasets \cite{z7,z8}.In summary, the evolution of blockchain-based IoT systems is characterized by the interplay between privacy enhancement and anonymous auditing within the blockchain framework. The goal is to create a secure, collaborative, and privacy-preserving environment where users can confidently engage with digital services. As these systems advance, security measures will become more sophisticated and integral to the decentralized web architecture\cite{a39,a40}.

\section{Integration of Blockchain-based IoT Systems with Emerging Technologies}

Blockchain-based IoT systems are integrating with cutting-edge technologies like Software-Defined Networking (SDN), Federated Learning (FL), and IoT devices to enhance security, privacy, and efficiency\cite{z6}.

\subsection{Software-Defined Networking (SDN) in Blockchain-based IoT Systems}

SDN's separation of control and data planes allows for flexible and efficient network management, which is crucial for the complex and dynamic nature of blockchain-based IoT systems. SDN's centralized control can effectively manage data flow, reduce latency, and improve performance in these systems. Furthermore, SDN's programmability extends to implementing sophisticated network functions directly into the infrastructure. This capability is vital in blockchain-based IoT systems, where security and reliability are of utmost importance. By integrating functions like load balancing and intrusion detection into network control, SDN provides a robust and secure foundation for decentralized applications and services. SDN also offers infrastructure abstraction, benefiting blockchain networks by allowing nodes to concentrate on core tasks like transaction verification without concern for underlying network conditions. This abstraction leads to more efficient resource usage and improved blockchain network performance\cite{a30,a31}.

In conclusion, SDN's unique features, including control-data plane separation, programmability, and infrastructure abstraction, make it ideal for managing the complexities of blockchain-based IoT systems. Its integration can optimize data pathways, enhance traffic management, and boost network performance, facilitating faster transaction verification and increasing overall network throughput.

 \begin{figure}[H]
    \centering
    \includegraphics[width=0.67\linewidth,keepaspectratio]{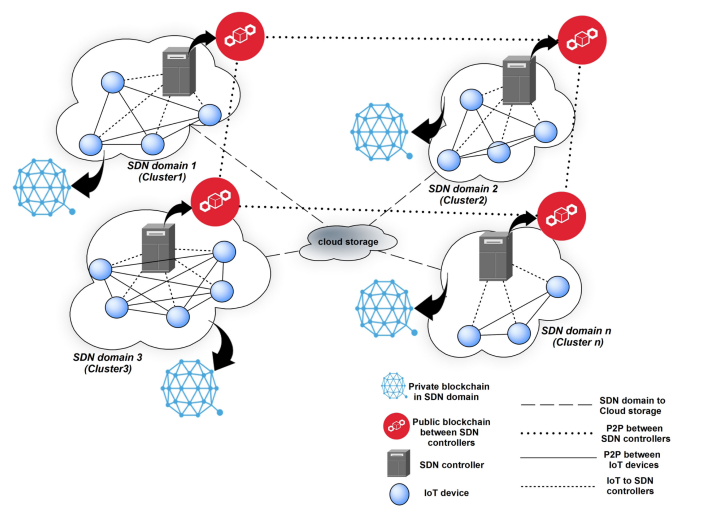}
    \caption{Workflow of A IoT SDN System}
    \label{fig11}
\end{figure}

\subsection{Federated Learning in Blockchain-based IoT Systems}

Federated Learning (FL) in blockchain-based IoT systems represents a significant evolution in how data is processed and machine learning models are trained. It aligns with the decentralized and privacy-centric principles of these systems. In traditional machine learning approaches, data from various sources is centralized for processing, posing risks of privacy breaches and data misuse. In contrast, FL in blockchain-based IoT systems keeps data on the device, sharing only model updates, thereby greatly reducing data leakage risks\cite{a37,a38}.The decentralized nature of FL offers several benefits for blockchain-based IoT systems:

\textbf{Privacy Preservation}: By allowing data to remain on the device, FL ensures privacy and security of user data. This approach aligns with the privacy requirements of many IoT applications, where sensitive data is generated and should be protected. Efficiency and Scalability: FL reduces the need for extensive data transmission, which is a significant challenge in large-scale IoT deployments. It enables faster model training and lowers communication costs, making it an efficient solution for machine learning in distributed IoT environments\cite{z9}. Personalized Models: Since FL trains models on local data, it captures unique patterns and characteristics of each node, leading to more personalized and accurate models. This enhances user experience and effectiveness of IoT applications.

\textbf{Integration with Blockchain}: Combining FL with blockchain technology enhances the security and transparency of the system. Blockchain can record model updates in an immutable ledger, ensuring the integrity of the FL process and increasing trust among participants. Resource Optimization: FL leverages the computational resources of IoT devices for model training, optimizing resource utilization across the network. This is especially beneficial in IoT environments where resources can be limited\cite{a32,a33}.

\subsection{Internet of Things (IoT) and Web 3.0 Integration}
The integration of IoT with Web 3.0 and blockchain opens up a plethora of opportunities for advanced, secure, and transparent applications:

Secure Data Logging: IoT-generated data can be securely logged on a blockchain, providing an immutable record that enhances transparency and trust in IoT applications\cite{a36,a37}.

\textbf{Smart Contract Automation:} IoT devices can interact with smart contracts on the blockchain, automating processes like supply chain management or automated payments based on IoT data inputs.

\textbf{Decentralized IoT Ecosystems}: Blockchain enables decentralized data exchanges among IoT devices, reducing reliance on central servers and enhancing system resilience.

\textbf{Enhanced Security and Privacy}: The combination of IoT and blockchain provides a dual layer of security—IoT for device-level security and blockchain for data integrity and privacy.

\textbf{Improved Device Management}: Blockchain can facilitate better device management and authentication in IoT networks, ensuring only authorized devices participate in the network\cite{z9,z10}.

\subsection{A Holistic Approach to Web 3.0 Security}
The integration of technologies like SDN, FL, and IoT into Web 3.0 exemplifies a holistic approach to creating a secure, efficient, and user-centric online ecosystem. SDN optimizes network performance, FL offers privacy-preserving machine learning, and IoT provides seamless connectivity, all under the secure and transparent framework of blockchain technology. This integration results in a robust Web 3.0 environment where security, privacy, and functionality are enhanced, paving the way for innovative applications and services in the decentralized digital world\cite{a34,a35}.

\section{Conclusion}
The integration of blockchain technology in IoT environments heralds a significant shift in the digital era, steering towards enhanced security, privacy, and user control. This convergence, while promising, introduces a range of unique security challenges, necessitating comprehensive solutions that are adaptive to the evolving nature of these technologies. This paper has illuminated these challenges, presenting a defined architecture of blockchain within IoT systems and proposing a security protocol specifically designed to augment privacy and enable anonymous auditing. Through our research, we've provided practical insights into the architecture and operational dynamics of blockchain-based IoT systems, with a concentrated effort on surmounting their intrinsic security challenges. The proposed security protocol, with its strong emphasis on privacy preservation and anonymous auditing, stands as a critical element in fortifying these systems. Implementing privacy-enhancing techniques and incorporating methods like Tor for anonymous auditing are integral to our model, showcasing a balanced approach to security and privacy.
Our comparative analysis has demonstrated the robustness and efficiency of our approach, setting it apart from existing methodologies. However, the continual evolution of blockchain and IoT technologies calls for an ongoing refinement and expansion of these security solutions. Future research will aim at enhancing the protocol's capabilities, ensuring it remains relevant and effective in the ever-advancing landscape of blockchain-based IoT systems.
As we progress in the digital transformation from traditional IoT to blockchain-integrated IoT, the importance of sophisticated security and privacy solutions becomes increasingly evident. Our research contributes significantly to this field, offering a foundation for continued exploration and development in creating secure, private, and reliable blockchain-based IoT environments.


\bibliographystyle{elsarticle-num}
\bibliography{References}

\end{document}